\documentstyle[12pt]{article}

\makeatletter
\def\eqnarray{\stepcounter{equation}\let\@currentlabel=\theequation
\global\@eqnswtrue
\global\@eqcnt\z@\tabskip\@centering\let\\=\@eqncr
$$\halign to \displaywidth\bgroup\@eqnsel\hskip\@centering
  $\displaystyle\tabskip\z@{##}$&\global\@eqcnt\@ne
  \hfil$\displaystyle{\hbox{}##\hbox{}}$\hfil
  &\global\@eqcnt\tw@ $\displaystyle\tabskip\z@
  {##}$\hfil\tabskip\@centering&\llap{##}\tabskip\z@\cr}
\@addtoreset{equation}{section}
  \def\theequation{\thesection.\arabic{equation}}
\makeatother

\advance \textheight by \topskip
\begin{document}

\title{Bosonized supersymmetry of anyons and
supersymmetric exotic particle on the non-commutative plane}

\author{
{\sf Peter A. Horv\'athy${}^a$}\footnote{ E-mails:
horvathy@univ-tours.fr}, {\sf Mikhail S.
Plyushchay${}^b$}\footnote{mplyushc@lauca.usach.cl},
{\sf Mauricio Valenzuela${}^b$}\footnote{mauricio.valenzuela@correo.usach.cl}\\
[4pt] {\small \it ${}^a$Laboratoire de Math\'ematiques et de
Physique Th\'eorique, Universit\'e de Tours,}\\
{\small \it Parc de Grandmont,
 F-37200 Tours, France}\\
{\small \it ${}^b$Departamento de F\'{\i}sica, Universidad de
Santiago de Chile, Casilla 307, Santiago 2, Chile} }

\date{
}

\maketitle

\begin{abstract}
A covariant set of linear differential field equations, describing
an  $N=1$ supersymmetric anyon system in (2+1)D, is proposed in
terms of Wigner's deformation of the bosonic Heisenberg algebra. The
non-relativistic ``Jackiw-Nair'' limit extracts the ordinary bosonic
and fermionic degrees of freedom from the Heisenberg-Wigner algebra.
It yields first-order, non-relativistic wave equations for a spinning particle on the
non-commutative plane that admits a  Galilean exotic planar
$N=1$ supersymmetry.
\end{abstract}

\section{Introduction}

Low dimensional quantum field theories may behave in a remarkable way~:
bosons and fermions can be equivalent \cite{Boson}. Similarly,
non-relativistic planar (or (2+1)D relativistic)
systems may exhibit a boson-fermion (or, more generally, a boson-anyon)
transmutation mechanism  \cite{Transmute}.

Yet another strange behaviour has been found
recently, when it was shown that, owing to these low dimensional peculiarities,
some purely bosonic quantum
mechanical systems 
may be supersymmetric \cite{SUSYf,bosSUSY}.
It appears that bosonization is related to  hidden nonlocal
structure. In bosonic systems with  hidden supersymmetry, in
particular, a nonlocal operator, namely reflection, plays the r\^ole
of a grading operator.

Quantum mechanics can be viewed as field theory in $0+1$ dimensions.
Then it is natural to ask  whether the bosonization can be extended
to  genuine  field theories
 which would be hence super-Poincar\'e symmetric.
The present paper
shows that this is indeed possible, namely in $2+1$ dimensions.

We start with observing that the
Fock space of the Heisenberg algebra $[a^-,a^+]=1$ can be viewed as
a direct sum of two irreducible infinite-dimensional unitary
representations of the Lorentz algebra $so(1,2)$, generated by
 operators which are quadratic in $a^\pm$. The two irreducible $so(1,2)$
representations are labeled by the Lorentz spin $1/4$ and $3/4$ (or
$-1/4$ and $-3/4$) and are realized in even and odd subspaces of the
Fock space. The latter are distinguished by the reflection operator
$R=(-1)^N=\cos \pi N$, $N=a^+a^-$. Note that $R$ is a nonlocal
operator (as it follows, e. g., from its infinite-series expansion).
These two $so(1,2)$ representations form an irreducible
representation of the $osp(1|2)$ superalgebra, generated by the even
(in the sense of Lie superalgebra) $so(1,2)$ generators and by the
odd generators $a^+$, $a^-$. They form an $so(1,2)$ spinor.  $R$
plays the r\^ole of a grading operator of the superalgebra.

Since the spins  in the two sectors differ by $1/2$, we have  a
priori all  ingredients needed to extend the construction to a
(2+1)D super-Poincar\'e symmetry. This can be done by treating
$a^\pm$ as describing internal (translation-invariant) spin degrees
of freedom of a supersymmetric system, which lives in
configuration space with cooordinates $x^\mu$. We introduce the fields
\begin{equation}
\Psi(x)=\sum_{n=0}^{\infty}\psi_n(x)\vert n \rangle, \qquad N\vert
n\rangle=n\vert n\rangle, \label{Psindec}
\end{equation}
 and  then postulate, ``\`a la Dirac'', a
covariant (vector) set of linear differential equations, which imply
the  Klein-Gordon equation  and Majorana-like equation as
integrability (consistency) conditions. These latters fix the two
Casimir operators of the (2+1)D Poincar\'e superalgebra, namely the
mass and the superspin, which fixes in turn the ``Poincar\'e
spins'' whose values, $1/4$ and $3/4$, are shifted  by one half.

Physical states are eigenstates with opposite $\pm1$ eigenvalues of
the reflection operator [playing, again, the r\^ole of a grading
operator of the Poincar\'e superalgebra], and will obey therefore a
relative Fermi statistics.

This construction provides us with  a bosonized  (2+1)D supersymmetry:
relative Fermi statistics is obtained without
introducing any fermionic degrees of freedom.
Everything is realized in terms of bosonic degrees of
freedom alone.

It is worth noting that Heisenberg structures leading to
``quartions'' (particles with $1/4$ spin) have been considered
before \cite{quartion,SorVol}. Infinite component Majorana-type
anyon equations were proposed in  \cite{MP1,JN1}.

Generalizing the ordinary Heisenberg algebra to the deformed
Heisenberg-Wigner algebra $ [a^-,a^+]=1+\nu R, $ where $\nu>-1$ is a
deformation parameter cf. (\ref{defHeis}) below, allows us to extend
the construction to a pair of two anyon fields with spins
$s=\frac{1}{4}(1+\nu)>0$ and $s+1/2>0 $ cf. (\ref{alphapm}) (or,
$s<0$, $s-1/2<0$ for the alternative realization of the Lorentz
generators), respectively.

For the  special values  $\nu=-(2k+1)$  the algebra has
finite-dimensional non-unitary representations \cite{MPRH}, see
the Discussion, Section \ref{discussion}. This algebra appeared
implicitly in Wigner's (1950) work \cite{wig}; its infinite-mode
generalization lead to the theoretical discovery of parastatistics
\cite{parab,OnuK}.

Below we mostly focus our attention to
the unitary case $\nu>-1$. Then the generators of the deformed
Heisenberg-Wigner algebra can be reinterpreted as ``entangled''
bosonic and fermionic degrees of freedom. Taking the special
nonrelativistic limit considered by Jackiw and Nair \cite{JaNa}
($s\rightarrow \infty$, $c\rightarrow \infty$, $s/c^2=\kappa=const$)
``extracts" these degrees of freedom, and produces a spinning
particle on the noncommutative plane. It carries an $N=1$
superextension of the ``exotic'' [i.e. two-parameter centrally
extended] Galilean symmetry.


Our investigations here generalize a previous attempt
\cite{HPPLB}, based on a similar construction \cite{Spinor}. In
\cite{HPPLB} a wave equation has indeed been proposed using a
spinor  operator, $Q_\alpha$ in  (2.7) of \cite{HPPLB}. When
restricted to the $+1$ eigenspace of the reflection operator $R$,
the Klein-Gordon and the Majorana equations were implied as
consistency conditions. Our wave equation could, therefore, be
considered as an infinite-component Majorana-type system,
analogous to those proposed before \cite{MP1,JN1}, which would
describe a relativistic anyon.

The subtle ``Jackiw-Nair'' non-relativistic limit \cite{JaNa} has
been designed by these authors to derive ``exotic particles''
(i.e., such that Galilean boosts don't commute) from  anyons. When
applied to the $R=+1$ sector of the model in \cite{HPPLB},
non-relativistic infinite-component Dirac-Majorana-type equations
are obtained with analogous properties. In particular, the
suitably defined Galilean boost operators ${\cal K}_i$ satisfy the
``exotic'' commutation relation
\begin{equation}
[{\cal K}_i,{\cal K}_j]=i\kappa\epsilon_{ij}. \label{exogalrel}
\end{equation}
These ${\cal K}_i$ take, furthermore, a natural form [(6.5) in
\cite{HPPLB}] in terms of some  Foldy-Wouthuysen-type coordinates
${\cal X}_i$ [(6.2) of \cite{HPPLB}], and the ``exotic'' relation
(\ref{exogalrel}) is reflected by their noncommutativity
\begin{equation}
[{\cal X}_i,{\cal X}_j]=i\theta\epsilon_{ij}, \qquad
\theta=\frac{\kappa}{m^2}. \label{NCcoordrel}
\end{equation}
In other words, the theory yields a noncommutative plane.

It has been puzzling whether the negative subspace $R=-1$ could
yield a supersymmetric partner of the bosonic model
 in \cite{HPPLB}. Easy calculation shows, however,
that the restriction of the wave equation of \cite{HPPLB} to the
subspace $R=-1$ is trivial. One of the sectors is hence lost.

In the present paper, this obstruction is overcome, though. Our
clue is to introduce another operator, namely (\ref{Lsusy3})
below, which, on the one hand, allows a nontrivial $R=-1$ sector
so that the $R=\pm1$ sectors become superpartners; on the other
hand, the restriction of the new operator to the $R=+1$ sector
yields a theory equivalent to the one in \cite{HPPLB}.


The paper is organized as follows. In Section 2, on the basis of
infinite-dimensional unitary representations of  Heisenberg-Wigner
algebra, we construct the anyon system admitting a bosonized
(2+1)D $N=1$ supersymmetry. In Section 3 a covariant set of linear
differential equations for this system is identified and their
general solution is discussed. In Section 4 we apply the
Jackiw-Nair non-relativistic limit and discuss the resulting
system of first order wave equations and its associated
supersymmetry. In Section 5 we discuss the appearance of
noncommutative coordinates in the theory. The last Section
includes concluding remarks, where, in particular, we briefly
discuss finite-dimensional representations of the deformed
Heisenberg-Wigner algebra.

\section{(2+1)D bosonized supersymmetry}

Consider the  deformed Heisenberg algebra  \cite{MPRH}
\begin{equation}\label{DHAR}
     \big[a^-,a^+\big]=1+\nu R,
     \qquad
     R^2=1,
     \qquad
     \{a^{\pm},R\}=0,
     \label{defHeis}
\end{equation}
where $\nu$ is a real deformation parameter. The usual
Heisenberg algebra $[a^-,a^+]=1$  is included as a particular case
$\nu=0$. The operator
\begin{equation}\label{Naa}
N=\frac{1}{2}\{a^+,a^-\}-\frac{1}{2}(\nu+1),\qquad [N,a^\pm]=\pm
a^\pm,
\end{equation}
is a number operator. Then, as in a non-deformed case, $R=(-1)^N=\cos\pi N. $

 The
$a^\pm$ satisfying the deformed Heisenberg algebra, satisfy also
trilinear commutation relations $[\{a^+,a^-\},a^\pm]=\pm 2a^\pm$,
see Eq. (\ref{Naa}), and vice versa \cite{mphid}. For $\nu=p-1,\,
p=1,2,\ldots$, the $a^\pm$ are creation-annihilation operators of a
single-mode paraboson of order $p$, characterized  by the additional
relation $a^-a^+\vert 0\rangle=p\vert 0\rangle$, where $\vert
0\rangle$ is a vacuum state, $a^-\vert 0\rangle=0$. In what follows
we refer to (\ref{DHAR}) as to Heisenberg-Wigner (HW) algebra.

For any $\nu>-1$, the algebra admits an infinite-dimensional unitary
representation realized on a Fock space; this is our main interest
here. The Fock space is spanned by the  states  $|n\rangle$,
$\langle n'|n\rangle=\delta_{n'n}$,
\begin{equation}\label{Fockn}
     N|n\rangle=n|n\rangle,\quad
     R|n\rangle=(-1)^n|n\rangle,\quad
     |n\rangle=C_{n}(a^+)^n|0\rangle,\quad
    n=0,1,\ldots,
\end{equation}
where $C_{n}$ is a normalization coefficient.

The Fock space (as well as any  finite-dimensional representation
of the HW algebra) is decomposed into even and odd subspaces
defined by $R|\psi\rangle_{\pm}=\pm|\psi\rangle_{\pm}$, which
correspond to $n$ being even or odd. Both for  finite and
infinite-dimensional representations, these
 subspaces carry irreducible
representations of the Lorentz algebra $so(1,2)$. The generators are
realized as quadratic operators,
\begin{equation}
     J_{0}=\frac{1}{4}\{a^+,a^-\},
     \qquad
     J_{\pm}\equiv J_1\pm iJ_2=\frac{1}{2}(a^{\pm})^2,
\label{a2}
\end{equation}
\begin{equation}\label{so12}
[J_\mu,J_\nu]=-i\epsilon_{\mu\nu\lambda}J^\lambda.
\end{equation}
[The antisymmetric  tensor is normalized by $\epsilon^{012}=1$;
the metric is $\eta_{\mu\nu}=diag (-1,1,1)$.] In the even and odd
subspaces the $so(1,2)$ Casimir  operator $J_\mu J^\mu$ takes
  the values $ J_{\mu}J^{\mu}=-\alpha_+(\alpha_+-1)$ and $
J_{\mu}J^{\mu}=-\alpha_-(\alpha_--1),$ where
\begin{equation}\label{alphapm}
    \alpha_+=\frac{1}{4}(1+\nu),\qquad
    \alpha_-=\alpha_++\frac{1}{2}.
\end{equation}
In the unitary (infinite dimensional) case  $\alpha_+>0$, and  the states $\vert n\rangle$
satisfy the relations
\begin{equation}\label{Jpos}
    J_0\vert 2k\rangle=(\alpha_++k)\vert 2k\rangle,\qquad J_0\vert
2k+1\rangle=(\alpha_-+k)\vert 2k+1\rangle,\quad k=0,1,\ldots,
\end{equation}
Every infinite-dimensional unitary representation of the deformed
Heisenberg algebra is therefore the  direct sum of two,
bounded-from-below, infinite-dimensional unitary representations of
the $so(1,2)$, $D^+_{\alpha_+} \oplus D^+_{\alpha_-}$ \cite{Bargm}
being (2+1)D analogs of the infinite-dimensional  unitary
representations of  $so(1,3)$  discovered by Majorana (1932)
\cite{Majorana}. The infinite-dimensional unitary representations
$D^-_{\alpha_+}$ and $D^-_{\alpha_-}$ bounded from above (necessary
to describe the states with negative spin values, see below) are
obtained via a simple sign change in (\ref{a2})
\begin{equation}
    J_0\to -\frac{1}{4}\{a^-,a^+\},\quad
    J_\pm\to-\frac{1}{2}(a^\mp)^2.\label{D-}
\end{equation}

The linear operators
\begin{equation}
{\cal L}_{1}=\frac{1}{\sqrt{2}}(a^++a^-),
    \qquad
{\cal L}_{2}=\frac{i}{\sqrt{2}}(a^+-a^-),
\end{equation}
extend the Lorentz algebra (\ref{so12}) to an $osp(1|2)$
superalgebra, in which $J_\mu$ and ${\cal L}_\alpha$ are even and
odd generators, respectively, and the grading operator is $R$,
$[R,J_\mu]=0,$ $\{R,{\cal L}_\alpha\}=0$, $R^2=1$. This superalgebra
is characterized, in addition to (\ref{so12}), by the
(anti)commutation relations
\begin{equation}\label{osp12}
\{{\cal L}_\alpha,{\cal L}_\beta\}=4i (J\gamma)_{\alpha\beta},\qquad
[J_{\mu},{\cal L}_{\alpha}]= \frac{1}{2}(\gamma_{\mu})_{\alpha}^{\
\beta}{\cal L}_{\beta}.
\end{equation}
Here the gamma-matrices are in the Majorana representation,
$$
(\gamma^0)_{\alpha}^{\ \beta}= -(\sigma^2)_{\alpha}^{\ \beta},
\qquad
(\gamma^1)_{\alpha}^{\ \beta}= i(\sigma^1)_{\alpha}^{\ \beta},
\qquad
(\gamma^2)_{\alpha}^{\ \beta}= i(\sigma^3)_{\alpha}^{\ \beta},
$$
where they satisfy relations
$$(\gamma_\mu)_\alpha{}^\rho(\gamma_\nu)_\rho{}^\beta=
    -\eta_{\mu\nu}\epsilon_{\alpha}{}^\beta+i\epsilon_{\mu\nu\lambda}
    (\gamma^\lambda)_\alpha{}^\beta,
    \qquad
\gamma^\mu_{\alpha\beta}=\gamma^\mu_{\beta\alpha},
\qquad
\gamma^{\mu\dagger}_{\alpha\beta}=-\gamma^\mu_{\alpha\beta}.
$$ The
antisymmetric tensor
$\epsilon^{\alpha\beta}=-\epsilon^{\beta\alpha}$
($\epsilon^{12}=1$),
provides us with a metric for the spinor indices,
$A^\alpha=\epsilon^{\alpha\beta}A_\beta$,
$A_\alpha=A^\beta\epsilon_{\beta\alpha}$. Note that $A^{\alpha}
B_\alpha=-A_\alpha B^\alpha $ for any $A_\alpha$ and $B_\alpha$.

 Every irreducible representation of the HW algebra is an
irreducible representation of the $osp(1|2)$ superalgebra, for which
its Casimir operator
\begin{equation}
 {\cal C}=J_\mu J^\mu-\frac{i}{8}{\cal L}^\alpha {\cal L}_\alpha
\end{equation}
 takes the fixed value ${\cal C}=\frac{1}{16}(1-\nu^2).$

 Since the $so(1,2)$ spin  has a relative
shift of one-half between the even and odd subspaces,
it would be natural to try to realize a (2+1)D
supersymmetry (in the sense of supersymmetric extension of the
Poincar\'e symmetry) using the HW algebra and its irreducible
representations.
For the purpose, we treat $a^\pm$ as internal, translation invariant
variables, augmented with independent space-time coordinates $x^\mu$
and conjugate momenta $p_\mu$, $[x_\mu,p_\nu]=i\eta_{\mu\nu}$,
$[x^\mu,a^\pm]=[p_\mu,a^\pm]=0$, and introduce a field
(\ref{Psindec}).
 Translations and complete
Lorentz  generators are identified as $P_\mu=p_\mu$ and
\begin{equation}
{\cal J}_\mu=-\epsilon_{\mu\nu\lambda}x^\nu p^\lambda+ J_\mu.
\label{caljmu}
\end{equation}
In accordance with Eqs. (\ref{caljmu}), (\ref{osp12}), the ${\cal
L}_{\alpha}$, $\alpha=1,2$, form a (2+1)D spinor. In its terms we
define an operator
\begin{equation}\label{QL}
    Q_\alpha=\frac{i}{\sqrt{2m(1+\nu)}}\,\left((p\gamma)_{\alpha}{}^\beta
    -mR
    \epsilon_\alpha{}^\beta\right){\cal
    L}_\beta.
\end{equation}
With the normalization depending on the mass parameter $m$, the
spinor operator $Q_\alpha$ has the dimension of a square root of
the space-time translation generator $P_\mu$, and can be
considered as a candidate for a supercharge operator of a
superextended (2+1)D Poincar\'e algebra, for which the operator
$R$ will play the r\^ole of a grading operator,
$$[R,P_\mu]=[R,{\cal J}]=0,
\qquad
\{R,Q_\alpha\}=0.
$$
 The space-time translation and Lorentz
generators,  together with the operator $Q_\alpha$ satisfy the
(anti)commutation relations
\begin{equation}\label{SuperPoi1}
    [P_\mu,P_\nu]=0,\qquad
    [{\cal J}_\mu, P_\nu]=-i\epsilon_{\mu\nu\lambda}P^\lambda,\qquad
    [{\cal J}_\mu,{\cal J}_\nu]=-i\epsilon_{\mu\nu\lambda}{\cal
    J}^\lambda,
\end{equation}
\begin{equation}\label{SuperPoi2}
    [P_\mu,Q_\alpha]=0,\qquad
    [{\cal J}_\mu,Q_\alpha]=\frac{1}{2}(\gamma_\mu)_\alpha{}^\beta
    Q_\beta,
\end{equation}
\begin{eqnarray}
    &\{Q_\alpha,Q_\beta\}=-2i(P\gamma)_{\alpha\beta}&\nonumber
    \\[4pt]
    &+\frac{2i}{m(1+\nu)}\left[(J\gamma)_{\alpha\beta}(p^2+m^2)-
    2(p\gamma)_{\alpha\beta}\left((pJ-m\alpha_+)\Pi_+
    +(pJ-m\alpha_-)\Pi_-\right)\right],&\label{SuperPoi3}
\end{eqnarray}
where  $\Pi_+=\frac{1}{2}(1+R)$ and $\Pi_-=\frac{1}{2}(1-R)$ are
projectors on the even and odd subspaces of the Fock space of the HW
algebra, respectively.
Decomposing our field as
\begin{equation}\label{Psi+-}
\Psi=\Psi_++\Psi_-,\qquad \Psi_\pm=\Pi_\pm\Psi,
\end{equation}
it is clear from Eq. (\ref{SuperPoi3}) that if the components
satisfy the Klein-Gordon and Majorana-like equations
\cite{Majorana,Majrev},
\begin{equation}\label{Psipm}
    (p^2+m^2)\Psi_\pm=0,\quad
    (pJ-m\alpha_+)\Psi_+=0,\quad
    (pJ-m\alpha_-)\Psi_-=0,
\end{equation}
then, on shell, the relation (\ref{SuperPoi3})  takes the form
\begin{equation}\label{QQpga}
    \{Q_\alpha,Q_\beta\}=-2i(P\gamma)_{\alpha\beta}.
\end{equation}
As a result, we obtain that the $N=1$ Poincar\'e superalgebra,
(\ref{SuperPoi1}), (\ref{SuperPoi2}), (\ref{QQpga}), is a symmetry
of the field system (\ref{Psipm}), realized without incorporating
fermionic degrees of freedom. Since $R\Psi_\pm=\pm\Psi_\pm$ and $R$
is  the grading operator of the Poincar\'e superalgebra, the fields
$\Psi_+$ and $\Psi_-$  describe massive  anyon states with spins
$s_+=\alpha_+$ and $s_-=s_++\frac{1}{2}$ and positive energy (see
Eq. (\ref{Jpos})), which carry relative Fermi statistics.

The operator
\begin{equation}\label{Cas2}
    {\cal C}=P^\mu{\cal J}_\mu+\frac{i}{8}Q^\alpha Q_\alpha
\end{equation}
is the Casimir operator of the superalgebra (\ref{SuperPoi1}),
(\ref{SuperPoi2}), (\ref{QQpga}). Then we find that on  states
satisfying equations (\ref{Psipm}) the superspin ${\cal
S}=\frac{1}{m}{\cal C}$ takes the value
\begin{equation}\label{Cal2val}
    {\cal S}=\frac{1}{2}(\alpha_++\alpha_-).
\end{equation}
This means that the field system (\ref{Psipm}) realizes an irreducible
representation of the  (2+1)D $N=1$ supersymmetry.

Note that in order to describe an $N=1$ supermultiplet with negative
spins $s_+=-\alpha_+$  and $s_-=-s_++\frac{1}{2}$ and positive
energies, it is sufficient to change a realization of the `internal
space' Lorentz generators $J_\mu$ according to Eq. (\ref{D-}) and
change $m\rightarrow -m$ in all relations except for the
normalization factor in the definition of the supercharge.

\section{Linear differential supersymmetric field equations}

Following  Dirac's idea \cite{Dirac,Majrev}, let us find linear
differential field equations for which the Klein-Gordon and
Majorana-like equations (\ref{Psipm})  appear as integrability
(consistency) conditions. This will provide us then with wave
equations for a spinning particle on the non-commutative plane,
which carries an irreducible representation of the $N=1$
superextended exotic Galilean symmetry. Consider indeed the vector
set of equations \cite{corpl},
\begin{equation}\label{lmu}
    V^{(\alpha)}_\mu\Psi^{(\alpha)}=0,\qquad
    V^{(\alpha)}_\mu=\alpha p_\mu-i\epsilon_{\mu\nu\lambda}p^\nu J^\lambda
    +m J_\mu,
\end{equation}
where we assume that the field $\Psi^{(\alpha)}$ carries an irreducible
representation of the $so(1,2)$ Lorentz group generated by the
translationally invariant operators $J_\mu$, and  characterized by the
relations
$$
J_\mu J^\mu\Psi=-\alpha(\alpha-1)\Psi^{(\alpha)}, \qquad
J_0\Psi^{(\alpha)}=(\alpha+n)\Psi^{(\alpha)}.
$$
Our representations can be unitary infinite-dimensional half-bounded
representations $D^+_\alpha$ realized on the even or odd subspaces
of the deformed Heisenberg algebra with parameter $\nu>-1$, as
described above. Alternatively, they can be finite-dimensional
non-unitary representations, realized by the same algebra with
parameter $\nu=-(2k+1)$.

Since the vector operator
$V^{(\alpha)}_\mu$ satisfies the commutation relation
\begin{equation}\label{VVcom}
    [V^{(\alpha)}_\mu,V^{(\alpha)}_\nu]=
    -i\epsilon_{\mu\nu\lambda}\left(\eta^{\lambda\rho}+(\alpha-1)^{-1}m^{-1}p^\lambda
    J^\rho\right)V^{(\alpha)}_\rho,
\end{equation}
the three equations (\ref{lmu}) form a consistent set.
On the other hand, this operator satisfies the relations
\begin{eqnarray}\label{pjl1}
    &J^\mu V^{(\alpha)}_\mu=(\alpha-1)(pJ-\alpha m),&
    \\[6pt]
    &p^\mu V^{(\alpha)}_\mu=\alpha (p^2+m^2)+ m(pJ-
    \alpha m),&\label{pjl2}
    \\[6pt]
    &i\epsilon^{\mu\nu\lambda}p_\nu J_\lambda V^{(\alpha)}_\mu=
    \alpha(\alpha-1)(p^2+m^2)+(pJ+ (\alpha
    -1)m)(pJ-\alpha m).&\label{pjl3}
\end{eqnarray}

Eq. (\ref{pjl1}) implies, in particular, that there is no
singularity in (\ref{VVcom}) at $\alpha=1$. {}From relations
(\ref{pjl1})-(\ref{pjl3}) we conclude that a field
$\Psi^{(\alpha)}$ which satisfies the system of vector equations
(\ref{lmu}) satisfies also the equations
$
 (p^2+m^2)\Psi^{(\alpha)}=0,
$
$
(pJ- m\alpha)\Psi^{(\alpha)}=0.
$
Hence $\Psi^{(\alpha)}$ carries an irreducible
representation of the (2+1)D Poincar\'e group, characterized by the
mass $m$, spin $s=\alpha$, and has positive energy $p^0>0$ if
the unitary representation $D^+_\alpha$ is chosen.

The operator $V^{(\alpha)}_\mu$ satisfies the identity
\begin{equation}\label{Rmu}
    W^\mu V^{(\alpha)}_\mu\equiv 0,\quad
    W_\mu= (\alpha-1)^2p_\mu
    -i(\alpha-1)\epsilon_{\mu\nu\lambda}p^\nu J^\lambda+
    (pJ)J_\mu.
\end{equation}
Only two components of the vector operator $V^{(\alpha)}_\mu$ are
therefore independent. All three components are necessary, however,
to have a manifestly covariant set of equations (\ref{lmu}) that
guarantees  the relativistic invariance of the theory \cite{SUSYf}.

In terms of  $J_\pm=J_1\pm iJ_2$, the $so(1,2)$ commutation
relations take a form $
    [J_0,J_\pm]=\pm J_\pm,
$
$
    [J_-,J_+]=2J_0.
$  For the bounded from below unitary representation
$D^+_\alpha$ we have therefore
\begin{eqnarray}
    &J_0\vert n)=(\alpha+n)\vert n),\quad
    J_+\vert n)=C^\alpha_n\vert n+1),\quad
    J_-\vert n)=C^\alpha_{n-1}\vert n-1),&\label{D+}
    \\[6pt]
    &C^\alpha_n=\sqrt{(2\alpha+n)(n+1)}.&\label{Cn}
\end{eqnarray}

Decomposing the field $\Psi^{(\alpha)}(x)$ into eigenstates of the
operator $J_0$, $\Psi^{(\alpha)}(x)=\sum_{n=0}^\infty \psi_n\vert
n)$, provides us with an equivalent, component form of the system
(\ref{lmu}),
\begin{eqnarray}
    &[\alpha(p^0-m)-nm]\psi_n+\frac{1}{2}\left[C^\alpha_np_+\psi_{n+1}-C^\alpha_{n-1}p_-\psi_{n-1}
    \right]=0,&\label{eqcomp0}
    \\[6pt]
    &-np_+\psi_n+C^\alpha_{n-1}(m-p^0)\psi_{n-1}=0,&\label{eqcomp+}
    \\[6pt]
    &(2\alpha+n)p_-\psi_n+C^\alpha_{n}(m+p^0)\psi_{n+1}=0.&\label{eqcomp-}
\end{eqnarray}

The dependence of the equations can also be seen by noting that a
suitable linear combination of
(\ref{eqcomp+}) and (\ref{eqcomp-}),
reproduces the  first equation (\ref{eqcomp0}).
 In the same way, suitable
linear combinations of any two equations from
(\ref{eqcomp0})--(\ref{eqcomp-}) show that every component
satisfies the Klein-Gordon equation. On the other hand, it also imply
the Majorana equation presented in component form,
$$
[p^0(\alpha +n)-\alpha
m]\psi_n+\frac{1}{2}[C^\alpha_np_+\psi_{n+1}+C^\alpha_{n-1}p_-\psi_{n-1}]=0.
$$

In the representation $D^+_\alpha$,  Eq. (\ref{eqcomp-})
allows us to express the component $\psi_{n+1}$ in terms of
$\psi_n$,
\begin{equation}\label{n+1n}
\psi_{n+1}=-\frac{2\alpha
+n}{C^\alpha_{n}}\,\frac{p_-}{p^0+m}\,\psi_n.
\end{equation}
Then substituting the relation (\ref{n+1n}) into Eq. (\ref{eqcomp+})
shows that every component satisfies the Klein-Gordon equation, and
repeated application of relation (\ref{eqcomp-}) allows us to
represent all  higher field components in terms of the lowest
one,
\begin{equation}\label{Psisol}
    \psi_n(p)=(-1)^n\sqrt{\frac{\Gamma(2\alpha+n)}{\Gamma(n+1)\Gamma(2\alpha)}}\,\left(\frac{p_-}{p^0+m}\right)^n
    \psi_0(p),
\end{equation}
\begin{equation}\label{psi0}
    \psi_0(p)=\delta\left(p^0-\sqrt{p_i^2+m^2}\right)f(p_i),
\end{equation}
where it is implied that we  switched to  momentum
representation.
Then some manipulations allow us to derive the equivalent set of
independent equations
\begin{eqnarray}
    &\sqrt{n+2\alpha}\,(m+p_0)\psi_n-\sqrt{n+1}\,p_+\psi_{n+1}=0,&\label{HP1}
    \\[6pt]
    &\sqrt{n+2\alpha}\,p_-\psi_n+\sqrt{n+1}\,(m-p_0)\psi_{n+1}=0,&\label{HP2}
\end{eqnarray}
which will be convenient for taking a special non-relativistic
limit to the system, see below.

Returning to our supersymmetric system, let us introduce a
notation for the even and odd states of the Fock space of the HW
algebra,
\begin{equation}
    \vert n)_+=\vert 2n\rangle,\quad
    \vert n)_-=\vert 2n+1\rangle,\quad
    R\vert n)_\pm=\pm \vert n)_\pm,\quad
    n=0,1,\ldots,
    \label{n+-def}
\end{equation}
and rewrite decomposition (\ref{Psindec}), (\ref{Psi+-}) in the form
\begin{equation}\label{Psinn+-}
    \Psi(x)=\Psi_+(x)+\Psi_-(x)=\sum_{n=0}^{\infty}\left(\psi^+_n\vert n)_+
+\psi^-_n\vert n)_-\right).
\end{equation}
Now, for every field $\Psi_+$ and $\Psi_-$, we postulate the vector
set of linear differential equations
 (\ref{lmu}) with the parameter $\alpha$ chosen  to be $\alpha_+$ and $\alpha_-$,
 respectively.
Taking the $so(1,2)$ generators $J_\mu$  as in Eq. (\ref{a2}), these
fields carry the $so(1,2)$  irreducible representations
$D^+_{\alpha_+}$ and $D^+_{\alpha_-}$, realized on
even and odd subspaces of the Fock space,
 respectively. It is clear that,
consistently with Eq. (\ref{Psipm}), we reproduce our supersymmetric
system. Therefore, the  covariant system of linear differential
equations for the field (\ref{Psinn+-}) we were looking for is
\begin{eqnarray}\label{Lsusy1}
    &V_\mu\Psi=0,&
    \\[6pt]
    &V_\mu=V^{(\alpha_+)}_\mu\Pi_+
    +V^{(\alpha_-)}_\mu\Pi_-=\frac{1}{4}\left(2+\nu-R\right) p_\mu-i\epsilon_{\mu\nu\lambda}p^\nu
J^\lambda
    +m J_\mu .&\label{Lsusy3}
\end{eqnarray}

As independent equations we can choose, again, the pair
 $V_+\Psi=0$, $V_-\Psi=0$, whose component form  can be
reduced to Eqs.  (\ref{HP1}) and (\ref{HP2}), in which the parameter
$\alpha$ takes the values $\alpha_+$ and $\alpha_-$ for
$\psi^+_n$ and $\psi^-_n$, respectively. The solutions of these equations for both
fields $\psi^+_n$ and $\psi^-_n$ are given by Eqs. (\ref{Psisol}),
(\ref{psi0}) with correspondingly chosen  value of $\alpha$. Note
that in the rest frame the solutions $\Psi_+$ and $\Psi_-$
are proportional to the Fock states $|0)_+=|0\rangle$ and
$|0)_-=|1\rangle$, i.e., the physical state with $p_i=0$ in the odd
subspace is  the first exited state of the Fock space.

Direct calculation shows that the spinor supercharge operator
(\ref{QL}) satisfies with the vector operator (\ref{Lsusy3}) the relation
\begin{equation}\label{VQ}
   [V_\mu,Q_\alpha]=(D_{\mu}{}^\nu)_\alpha V_\nu,
\end{equation}
where $(D_{\mu}{}^\nu)_\alpha$ is some operator, whose explicit form
is not needed for us here. Relation (\ref{VQ}) means that the
supercharge (as well as the even generators of the (2+1)D superalgebra)
is a symmetry generator: acting on a physical state satisfying Eq.
(\ref{Lsusy1}), it produces another physical state.

Now the complex linear combinations
\begin{equation}
Q_\pm=Q_1\mp iQ_2
\end{equation}
are eigenstates of the rotation operator,
\begin{equation}\label{JQpm}
    [{\cal J}_0,Q_\pm]=\pm \frac{1}{2}Q_\pm.
\end{equation}
In terms of generators of the HW algebra, their explicit form is
\begin{eqnarray}
     &Q_-=\frac{i}{\sqrt{m(1+\nu)}}\left[a^-(mR+p_0)-a^+p_-\right],&\label{Q-}
     \\[4pt]
      &Q_+=\frac{i}{\sqrt{m(1+\nu)}}\left[a^+(mR-p_0)+a^-p_+\right].&\label{Q+}
\end{eqnarray}
In the rest frame system (where  $p^0=m$ for physical states), these
reduce to
\begin{equation}\label{Qp=0}
    Q_-(p_i=0)\approx -2i\sqrt{\frac{m}{1+\nu}}\, a^-\,\Pi_-,\qquad
    Q_+(p_i=0)\approx 2i\sqrt{\frac{m}{1+\nu}}\, a^+\,\Pi_+.
\end{equation}
{}From Eq. (\ref{Qp=0}) it is clear that the operator $Q_-$  transforms
the rest frame physical state $|0)_-=|1\rangle$ into the the
physical state $|0)_+=|0\rangle$ and annihilates the latter, while the
operator $Q_+$ acts in the opposite way.
In other words,  the action of the mutually
conjugate operators $Q_+$ and $Q_-$ corresponds to the action of
ordinary nilpotent supercharges in the system with $N=1$
supersymmetry.

In conclusion of this section we note that in \cite{HPPLB} instead
of vector set of dependent equations, the spinor set of
independent equations \cite{Spinor}
\begin{equation}\label{Salpha}
    \tilde{Q}_\alpha\Psi=0,\qquad
    \tilde{Q}_\alpha=\left(R(p\gamma)_\alpha{}^\beta+m\epsilon_\alpha{}^\beta\right){\cal
    L}_\beta
\end{equation}
was used [in \cite{HPPLB}, the spinor operator is denoted by
$Q_\alpha$] . The relation between vector and spinor operators is
given by
\begin{equation}\label{VQal}
    {\cal L}_\alpha(\gamma_\mu)^{\alpha\beta}R\tilde{Q}_\beta=
    -4i\left(V^{(\alpha_+)}_\mu\Pi_+-\left(V^{(\alpha_-)}_\mu+2i\epsilon_{\mu\nu\lambda}p^\nu
    J^\lambda-p_\mu\right)\Pi_-\right).
\end{equation}
This relation shows that both systems are closely related. In the
even subspace the two systems are indeed equivalent. This is
readily seen using the component  forms (\ref{eqcomp0}),
(\ref{eqcomp+}), (\ref{eqcomp-}) and (\ref{HP1})--(\ref{HP2}),
respectively.

In the odd subspace, the  spinor set of equations (\ref{Salpha})
has only trivial  solution $\Psi_-=0$. The new, vector set of
equations (\ref{Lsusy1}), (\ref{Lsusy3}), however, does have  a
nontrivial solution (\ref{Psisol}), (\ref{psi0}) with shifted
value of spin $\alpha=\alpha_-=\alpha_++\frac{1}{2}$.

\section{Jackiw-Nair nonrelativistic limit}

Let us now consider the special non-relativistic ``Jackiw-Nair'' (JN) limit \cite{JaNa}
of our relativistic supersymmetric system,
\begin{equation}\label{JN}
    c\rightarrow \infty,\qquad
    s\rightarrow \infty,\qquad
     \frac{s}{c^2}=\kappa.
\end{equation}
 For the purpose
we need some details on the Fock representations of the HW
algebra \cite{MPRH}. In the unitary case  $\nu>-1$ the normalization
coefficients in Eq. (\ref{Fockn}) are
$$C_n=([n]_\nu!)^{-1/2},
\qquad
[n]_\nu!=\prod_{l=1}^n[l]_\nu,
\qquad
[l]_\nu=l+\frac{1}{2}\left(1-(l)^l\right)\nu.
$$
 Then, consistently
with these relations and also with Eq. (\ref{DHAR}), we have
\begin{equation}
    a^+\vert n\rangle =\sqrt{[n+1]_\nu}\,\vert n+1\rangle,\quad
    a^-\vert n\rangle =\sqrt{[n]_\nu}\,\vert n-1\rangle.
\label{aak}
\end{equation}
Eq. (\ref{aak}) and definition (\ref{n+-def}) yield the relations
\begin{eqnarray}
    &a^+\vert n)_+=\sqrt{2(n+2\alpha_+)}\, \vert n)_-,\quad
    a^+\vert n)_-=\sqrt{2(n+1)}\,\vert n+1)_+\,,&\label{aaevod+}
    \\[6pt]
    &a^-\vert n)_+=\sqrt{2n}\,\vert n-1)_-,\quad
    a^-\vert n)_-=\sqrt{2(n+2\alpha_+)}\,\vert n)_+\,.&\label{aaevod-}
\end{eqnarray}

Let us introduce the ordinary bosonic operators $b^\pm$,
$$
[b^-,b^+]=1,
\qquad
N_b| n\rangle_b=n| n\rangle_b,
\qquad
N_b=b^+b^-.
$$
and represent the states of the even and odd subspaces as
\begin{equation}\label{Psi+-b}
    |n)_+=\left(
\begin{array}{c}
  1 \\
  0 \\
\end{array}
\right)\otimes |n>_b,
\quad |n)_-=\left(
\begin{array}{c}
  0 \\
  1 \\
\end{array}
\right)\otimes |n>_b.
\end{equation}
Then $R=\tau_3\otimes 1$, where, in order to distinguish it from
that  in the gamma-matrices, the diagonal Pauli
matrix  is denoted by $\tau_3$. By (\ref{aaevod+}) and
(\ref{aaevod-}), in the representation (\ref{Psi+-b}), the operators
$a^\pm $ are
\begin{equation}\label{asig}
    a^+=\sqrt{2}\left(\tau_+\otimes b^++\tau_-\otimes
    \sqrt{N_b+2\alpha_+}\right),\quad
    a^-=\sqrt{2}\left(\tau_-\otimes b^-+\tau_+\otimes
    \sqrt{N_b+2\alpha_+}\right),
\end{equation}
where $\tau_\pm=\frac{1}{2}(\tau_1\pm i\tau_2)$. Using
 $\{\tau_+,\tau_-\}=1$, $[\tau_+,\tau_-]=\tau_3$, we have
here $[a^-,a^+]=1+\nu\tau_3\otimes 1$.

In this representation, the Lorentz generators (\ref{a2}) are
\begin{eqnarray}
    &J_0=1\otimes (N_b+\alpha_+)+\frac{1}{4}(1-\tau_3)\otimes 1,&\label{J0tau}
    \\[6pt]
    &J_+=\frac{1}{2}(1+\tau_3)\otimes b^+\sqrt{N_b+2\alpha_+}
    +\frac{1}{2}(1-\tau_3)\otimes b^+\sqrt{N_b+2\alpha_++1}\,
    ,&\label{J+tau}\\
    &J_-=\frac{1}{2}(1+\tau_3)\otimes \sqrt{N_b+2\alpha_+}\, b^-
    +\frac{1}{2}(1-\tau_3)\otimes \sqrt{N_b+2\alpha_++1}\,b^-\,
    .&\label{J-tau}
\end{eqnarray}
Omitting the direct product symbol, (\ref{J+tau}) and
(\ref{J-tau})  can be written alternatively
\begin{eqnarray}
    &J_+=b^+\sqrt{N_b+2\alpha_++\frac{1}{2}(1-\tau_3)}\, ,\qquad
    J_-=\sqrt{N_b+2\alpha_++\frac{1}{2}(1-\tau_3)}\, b^-\,.&
\end{eqnarray}

Restoring  the dependence on the velocity of light,
 $c$, via the change $m\rightarrow mc$, the linear
differential equations (\ref{HP1}), (\ref{HP2}) for both fields
$\Psi_+$ and $\Psi_-$ are represented  in  operator form as
\begin{eqnarray}
    &\left((mc+p_0)\sqrt{2J_0-N_b}-p_+b^-\right)\Psi=0,&\label{fieldtau1}
    \\[6pt]
    &\left(p_-\sqrt{2J_0-N_b}+(mc-p_0)b^-\right)\Psi=0.&\label{fieldtau2}
\end{eqnarray}
These two equations are nothing else as the two independent equations
$V_+\Psi=0$ and $V_-\Psi=0$ from the covariant set (\ref{Lsusy1}).

Now let us consider a JN limit (\ref{JN}) with
$s=\alpha_+=\frac{1}{4}(1+\nu)$. Then, from 
(\ref{asig}), we find that
\begin{equation}\label{anu}
    \frac{a^+}{\nu^{1/2}}\rightarrow \tau_-\otimes 1,\qquad
    \frac{a^-}{\nu^{1/2}}\rightarrow \tau_+\otimes 1,
\end{equation}
i.e., the appropriately rescaled creation-annihilation operators are
transformed, in the limit $\nu\rightarrow \infty$, into fermion
operators (see also \cite{MPRH}). We have also
\begin{equation}\label{JpmJN}
    \frac{J_\pm}{c}\rightarrow \sqrt{2\kappa}\, b^\pm.
\end{equation}

On the other hand, in the JN limit, the energy and angular momentum
diverge. The ``renormalized" angular momentum \footnote{ There is an
arbitrariness in this procedure up to an additive constant
\cite{DHSpin}; here the constant is chosen in such a way that
non-relativistic spin will take zero value for the state
corresponding to relativistic state with spin $s=\alpha_+$.},
$J_0-\alpha_+$, becomes, in the JN limit, the operator
$N_b+\frac{1}{4}(1-\tau_3)$. Following \cite{HPPLB}, we define the
`velocity' operators
$$
v_\pm=-\sqrt{\frac{2}{\kappa}}\, b^\pm,
\qquad
[v_-,v_+]=2\kappa^{-1}.
$$
  The Galilei boosts are defined as ${\cal
K}_i=-\frac{1}{c}\epsilon_{ij}{\cal J}_j$. Then, for the (total) rotation
and Galilei boosts, we get, in the JN limit,
\begin{equation}\label{JK}
    {\cal J}=\epsilon_{ij}x_ip_j+\frac{1}{2}\kappa
    v_+v_-+\frac{1}{4}(1-\tau_3),\qquad
    {\cal K}_i=mx_i-tp_i+\kappa\epsilon_{ij}v_j.
\end{equation}
These operators will span an ``exotic'' Galilei algebra, see
(\ref{eG1}), (\ref{eG2}).  Eqs. (\ref{anu}) and (\ref{JpmJN}) show
that the JN limit extracts from the parabosonic-like operators
$a^\pm$ the ordinary bosonic and fermionic degrees of freedom.

In order to identify  the Hamiltonian  and the wave equations of the
JN limit of our supersymmetric system, we put in Eqs.
(\ref{fieldtau1}), (\ref{fieldtau2}) $p_0=-ic^{-1}\partial_t$ and
$\Psi=e^{-imc^2t}\Phi$, and apply (\ref{JN}). This  results in the
equations
\begin{eqnarray}
    &\left(i\partial_t-\frac{1}{2}p_+v_-\right)\Phi=0,\qquad
    (p_--mv_-)\Phi=0,&\label{phi2}
\end{eqnarray}
where  $\Phi$ is a two-component field  on which the spin matrices
$\tau$ act. Every such a component is decomposed in Fock space
states of the bosonic operators $b^\pm$, $\Phi_\pm=\sum_{n=0}^\infty
\phi^\pm_n|n\rangle_b$. The second equation in  (\ref{phi2}) is a
constraint, allowing, as in a relativistic case, to present all
higher components in terms of the lowest ones,
$$
\phi^\pm_n=(-1)^n\frac{1}{\sqrt{n!}}\left(\sqrt{\frac{\kappa}{2}}\,\frac{p_-}{m}\right)^n\phi^\pm_0.
$$
The substitution of the second equation from (\ref{phi2}) into the
first one shows that every component $\phi^\pm_n$ satisfies the
Schr\"odinger equation of a free non-relativistic particle. Adding
 the first equation to the second one multiplied from the left
by $-\frac{1}{2}v_+$ allows us to identify finally the Hermitian operator
\begin{equation}\label{calH}
    {\cal H}=p_i v_i -\frac{1}{2}v_+v_-,
\end{equation}
$v_\pm=v_1\pm i v_2$, as a  Hamiltonian of the nonrelativistic
system. Note that ${\cal H}$ is linear in the momentum. The wave
equations and the Hamiltonian coincide hence with those  \cite{HPPLB}
corresponding to the spinless exotic particle  on the
non-commutative plane \cite{DH}.

Now, let us identify the JN limit of the supercharge and
corresponding superalgebra. Restoring the light speed in $Q_\pm$ 
then defining
$$
{\cal Q}_\pm=c^{-1/2}Q_\pm
$$
and taking into account Eq. (\ref{anu}), we find that in the J-N
limit the supercharges reduce to
\begin{equation}\label{calQ}
    {\cal Q}_-=-2i\sqrt{m} \, \tau_+,\qquad
    {\cal Q}_+=2i\sqrt{m}\, \tau_-.
\end{equation}

The  system has a superextended exotic Galilei
symmetry, whose bosonic part,
\begin{eqnarray}
    &[{\cal K}_i,p_j]=im\delta_{ij},\quad
    [{\cal K}_i,{\cal K}_j]=-i\kappa\epsilon_{ij},&\label{eG1}
    \\[6pt]
    &[{\cal K}_i,{\cal H}]=ip_i,\quad
    [{\cal J},p_i]=i\epsilon_{ij}p_j,\quad
    [{\cal J},{\cal K}_i]=i\epsilon_{ij}{\cal K}_j,&\label{eG2}
\end{eqnarray}
is augmented by the  (anti)commutation relations involving the
supercharges,
\begin{eqnarray}
    &[{\cal J},{\cal Q}_\pm]=\pm \frac{1}{2}{\cal Q}_\pm,\quad
     \{{\cal Q}_+,{\cal Q}_-\}=4m,&\label{JQQ1}
     \\[6pt]
     & [{\cal K}_i,{\cal Q}_\pm]=[P_i,{\cal Q}_\pm]=[{\cal H},{\cal
    Q}_\pm]=0,\quad {\cal Q}_\pm^2=0.&\label{JQQ2}
\end{eqnarray}

The Casimir operators of the superextended exotic Galilei
symmetry are
\begin{equation}\label{CC}
    {\cal C}_1=p_i^2-2m{\cal H},\qquad
    {\cal C}_2={\cal J}-\epsilon_{ij}{\cal K}_i p_j +\kappa
    {\cal H}
    -\frac{1}{16}[{\cal Q}_+,{\cal Q}_-],
\end{equation}
where ${\cal C}_2$ is the JN limit of the superspin ${\cal
S}=m^{-1}{\cal C}$ with ${\cal C}$  given by Eq. (\ref{Cas2}). On
states satisfying equations (\ref{phi2}), these Casimir
operators take the values ${\cal C}_1=0$ and ${\cal
C}_2=\frac{1}{4}$.

The $N=1$ supersymmetric extension  (\ref{eG1}), (\ref{eG2}),
(\ref{JQQ1}), (\ref{JQQ2}) of the exotic Galilei symmetry was
discussed in \cite{SUSYGal,SUSYGal2}.

\section{Noncommutative coordinates}

Here we discuss how noncommutative coordinates appear in the
theory.

 In the relativistic theory, we start with the usual, commuting
space-time coordinates $x_\mu$. The physical subspace of the
system is given by the system of equations (\ref{Psipm}).  Taking
into account the first, Klein-Gordon equation, the two other,
Majorana type equations can be unified into a single equation,
namely into
\begin{equation}\label{MajMod}
   \chi_s\Psi\equiv \left(\frac{pJ}{\sqrt{-p^2}}- \left(\alpha_+\Pi_+
    +
    \alpha_-\Pi_-\right)\right)\Psi=0.
\end{equation}
The initial space-time coordinates $x_\mu$ are not observable with
respect to the superspin equation (\ref{MajMod}): since
$[x_\mu,\chi_s]\neq 0$, acting on a physical state satisfying Eq.
(\ref{MajMod}), the operator $x_\mu$ produces a state which does
not belong to the physical subspace. Let us define instead
\cite{CorMP}
\begin{equation}\label{X*}
    {X}_\mu=x_\mu+\frac{1}{p^2}\epsilon_{\mu\nu\lambda}p^\nu
    J^\lambda.
\end{equation}
These modified coordinates satisfy the relation
\begin{equation}\label{Xchi}
    [{X}_\mu,\chi_s]=0,
\end{equation}
and so, are observable operators. ${ X}_\mu$ is a vector operator
with respect to the (2+1)D Lorentz transformations generated by
(\ref{caljmu}). However, unlike the initial coordinates, the
coordinates ${X}_\mu$ are not commuting,
\begin{equation}\label{XXnon}
    [{ X}_\mu,{ X}_\nu]\approx -i\left(\alpha_+\Pi_+
    +\alpha_-\Pi_-\right)\epsilon_{\mu\nu\lambda}\frac{p^\lambda}{(-p^2)^{3/2}},
\end{equation}
where the symbol $\approx$ means ``on the surface defined by
equation (\ref{MajMod})". Therefore, the physical (observable)
coordinates are non-commuting. They are analogous to the
Foldy-Wouthuysen coordinates of the Dirac particle
\cite{FW}.

In the non-relativistic Jackiw-Nair limit the space part of the
coordinates (\ref{X*}) is transformed into \cite{HPPLB}
\begin{equation}\label{xV}
    {\cal X}_i=x_i+\frac{\kappa}{m}\epsilon_{ij}V_j+\frac{\theta}{2}\epsilon_{ij}p_j,
\end{equation}
where
\begin{equation}\label{Vivi}
    V_i=v_i-\frac{1}{m}p_i
\end{equation}
and $\theta=\kappa/m^2$. The operator in the second equation in
(\ref{phi2}) is a complex linear combination of the operators
$V_i$, $i=1,2$, that behaves as an annihilation operator (an
operator with nontrivial kernel). The coordinates (\ref{xV}), like
the initial coordinates $x_i$, form a 2D vector, which is a
covariant object with respect to Galilei boosts generated by the
${\cal K}_i$ from (\ref{JK}) (${\cal X}_i$ commutes with ${\cal
K}_j$ in the same way as $x_i$ does).

The initial coordinates $x_i$ do  not commute with the operator
$-mV_-$  appearing in the second equation from (\ref{phi2}), and
are subject to a Zitterbewegung-like motion under the evolution
generated by the first order Hamiltonian (\ref{calH})
\cite{HPPLB},
\begin{equation}\label{dotxiv}
\dot{x}_i=v_i,\qquad
\dot{v}_i=\frac{m}{\kappa}\epsilon_{ij}V_j.
\end{equation}
Unlike $x_i$, the coordinates ${\cal X}_i$ commute with the
constraint operator $-mV_-$, and so are observable operators. They
have the usual (Zitterbewegung-free) evolution of the coordinates of a free non-relativistic particle,
\begin{equation}\label{dotcalX}
\dot{\cal X}_i=\frac{1}{m}p_i.
\end{equation}
However, the components of the coordinate (\ref{xV}) do not
commute,
\begin{equation}\label{XXtheta}
    [{\cal X}_i,{\cal X}_j]=i\theta\epsilon_{ij},
\end{equation}
and describe therefore a non-commutative plane \cite{HP1,HPPLB}.

\section{Concluding remarks}\label{discussion}

In the non-supersymmetric case, the system of dependent vector
equations (\ref{lmu}) was realized originally on the basis of an
irreducible representation of the (2+1)D Lorentz algebra
\cite{corpl}. It can be replaced by an independent covariant
spinor set of equations (\ref{Salpha}) constructed in terms of an
irreducible Fock space representation of the deformed
Heisenberg-Wigner algebra. The corresponding non-supersymmetric
anyon system is described by a field  with nontrivial even part
$\Psi_+$ alone, living in the even subspace of the Fock
representation. In the present, supersymmetric, case, however, no
similar substitution of the vector for spinor set of equations
seems to exist for both, even and odd, fields $\Psi_+$ and
$\Psi_-$. This may well explain a failure of a previous attempt to
construct a bososonized (2+1)D supersymmetry \cite{SUSYfail},
where
 minimal, spinor set of equations has been searched for.
 Here, like in \cite{corpl}, our basic set of equations (\ref{Lsusy1}), (\ref{Lsusy3}) is
 vector, but it is realized on an  irreducible
Fock space representation of the deformed HW algebra, which is an
irreducible representation of the $osp(1|2)$ superalgebra. By
using as a grading operator the reflection operator of the
deformed HW algebra, we obtained a (2+1)D supersymmetric system
avoiding introduction of fermionic Fock space.

The equations which correspond to finite-dimensional non-unitary
representations of HW algebra  behave similarly to the anyonic
case. There is, however, an essential difference in the
realization of supersymmetry. Finite-dimensional representations
provide us with parafermion-like degrees of freedom described by
the $a^\pm$, see Ref. \cite{MPRH}. In this case we have ordinary,
not bosonized  (2+1)D supersymmetry, realized  on a system of
ordinary integer or half-integer spin fields. For $\nu=-(2k+1)$,
$k=2,3,\ldots$, the eigenvalues of the operator $J_0$ can take
both signs, yielding two independent solutions of the system of
linear differential equations (\ref{Lsusy1}), namely  with either
positive or negative energy. For the positive energy solutions the
components $\psi^+_n$, $n=0,1,\ldots, k$, of a spin
$s_+=-\frac{k}{2}$ field and the components $\psi^-_{n'}$,
$n'=0,1,k-1$, of a spin $s_-=-\frac{k-1}{2}$ field can be
expressed in terms of the components $\psi^+_0$ and $\psi^-_0$,
via the equation $V_-\Psi=0$, by relations similar to
(\ref{Psisol}). For negative energy solutions these are
represented via  $V_+\Psi=0$ in terms of
 $\psi^+_k$ and $\psi^-_{k-1}$.

For example, $\nu=-5$ yields a  spin $1/2$ fermion plus an $s=1$
(topologically massive gauge) vector field. The case $\nu=-3$ (i.e.
$k=1$) is a special one. Now $s_-=\alpha_-=0$, and the corresponding
$so(1,2)$ Lorentz generators $J_\mu$ in the odd subspace are
trivial. Hence, the operators $V^{(\alpha_-)}_\mu$ as well as the
spin fixing linear differential operator $(PJ-\alpha_-m)$ disappear.
For a scalar field $\Psi_-$ our construction breaks down. We still
get  a supermultiplet of an $s=0$  scalar plus a spin $1/2$ fermion
field; for the spin-zero superpartner, however, the dynamics is
quadratic Klein-Gordon equation, rather than a first-order one.

Note that the even generators $P_\mu$ and ${\cal J}_\mu$  of the
bosonized (2+1)D supersymmetry of the anyon system are local in
internal space, associated with the operators $a^\pm$. The odd
supercharge $Q_\alpha$ is, however, nonlocal due to the presence of
 the operator $R$. This is similar to what we have
for  purely bosonic quantum mechanical systems with hidden
supersymmetry \cite{bosSUSY}, where the Hamiltonian is local but odd
supercharges are nonlocal operators.

The application of the Jackiw-Nair non-relativistic limit to our
theory yields  wave equations describing a spinning particle on
the noncommutative plane,  and that the (2+1)D Poincar\'e
supersymmetry reduces, in this limit, to Galilean exotic $N=1$
supersymmetry. The latter is characterized by the supercharges
being square roots of one of the central charges, namely of the
mass, $m$. One could
 expect that if the  relativistic supersymmetric anyon proposed here
 is extended by adding ordinary fermion degrees of
freedom (for instance, within the framework of a superfield approach
\cite{SorVol}), the resulting system would possess $N=2$
supersymmetry, half of which would be   bosonized. Then subsequent
application of the Jackiw-Nair limit
should produce a system possessing a Galilean exotic $N=2$
supersymmetry with new supercharges which would be the square roots of a
Hamiltonian \cite{SUSYGal}.

In conclusion, we have shown that bosonized supersymmetry can be realized in a
(2+1)D system of anyons, constructed on the basis of
infinite-dimensional unitary irreducible representations of the
Heisenberg-Wigner algebra. Any such a representation carries two
irreducible unitary representations of the $so(1,2)$ Lorentz group,
whose spins are shifted relatively by one-half. These
infinite-dimensional representations of the (2+1)D Lorentz group are
analogs of the infinite-dimensional unitary representations of the
(3+1)D Lorentz group, discovered by Majorana in his celebrated 1932
paper \cite{Majorana,Majrev}.
It might be possible  to generalize our construction  of the bosonized
supersymmetry here to  (3+1)-dimensional field theory.
This problem will be studied elsewhere.

\vskip 0.4cm\noindent {\bf Acknowledgements}.  MP is indebted to
the Laboratoire de Math\'ematiques et de Physique Th\'eorique of
Tours University for a hospitality extended to him. PAH thanks
Departamento de F\'{\i}sica of Universidad de Santiago de Chile
for hospitality during his visit. This work was supported in part
by FONDECYT (projects 1050001 and 7060057) and by MECESUP USA0108.

\end{document}